\documentstyle[12pt,leqno]{article}
\newcommand{\proof}{{\bf Proof. \ }}
\newcommand{\qed}{\hfill $\Box$ \\}

\font\msbm=msbm10 at 12pt
\newcommand{\ZZ}{\mbox{\msbm Z}}

\def \Z {{\ZZ}}
\def \z {{\ZZ}}
\def \C {{\cal C}}

\def \v {{\bf v}}

\def \B {{\cal B}}

\def \v {{\bf v}}

\def \v {{\bf v}}

\newtheorem{theorem}{Theorem}
\newtheorem{lemma}{Lemma}
\newtheorem{proposition}{Proposition}
\newtheorem{remark}{Remark}
\newtheorem{example}{Example}

\newcommand{\be} {\begin{equation}}
\newcommand{\ee} {\end{equation}}

\begin{document}

\title{On Modular Gray Map}
\author{Manish K Gupta
\thanks{A part of this work was presented at the Special Session on Algebraic Coding Theory, Central Section Meeting \#995, American Mathematical Society, Athens,OH, 26-27 March 2004.}\\
Dhirubhai Ambani Institute of Information and Communication Technology,\\
Post Bag Number 4 , Near Indroda Circle\\
Gandhinagar,   382007,
India\\
{Email:mankg@guptalab.org}
}

\maketitle


\vspace*{0.5cm}

\begin{abstract}
This paper introduces an isometry between the modular rings $\Z_{2^s}$ and $\Z_{2^{s-1}}$ with respect to the homogeneous weights.  Certain product of these maps gives Carlet's generalised Gray map and also Vega's Gray map. For $s=2$ this reduces to popular Gray map. Several interesting properties of these maps are studied.  Towards the end we list several interesting problems to work on.
\end{abstract}
\vspace*{0.5cm}

\vspace{0.5cm}

\newpage

\section{Introduction}

In $19^{th}$ century a cyclic-permuted binary code was invented
by a french engineer {\em Emile Baudot} (now often called {\em
Gray code} as it was patented by {\em Frank Gray} on $17.03.1953$
\cite{gr53}).  This has represented a major advancement in telegraphy.
 After that it has witnessed various practical applications other
than solving puzzles such as Tower of Hanoi and the Brain \cite{web1}.
It is very recent when it is used as an isometry between $\Z_4$ and
$\Z_2^{2}$ with respect to the Lee and Hamming distances to solve
a $30$ year old coding theory puzzle \cite{hkcss94}. This isometry can
be extended component-wise between $\Z_4^{n}$ and $\Z_2^{2n}$ and
is usually known as the {\em Gray map}. Due to an isometry the
Gray map has been used extensively to construct good quality
binary linear and nonlinear codes \cite{wan97,gu01,hkcss94}.  Also the $\Z_4$-
linearity of various binary codes of even length has been
inevstigated.  In \cite{sa99}, Ana $S\breve{a}l\breve{a}gean$-
Mandache has shown that except for the well-known case $p = s
=2,$ it is not possible to construct a weight function on
$\;\Z_{p^{s}}$ for which $\Z_{p^{s}}$ is isometric to $\Z_p^{s}$
with the Hamming metric. A similar result for the Lee metric was established 
in \cite{co97} by S. R. Costa et al. However there exist an isometry
between $\Z_{2^{s}} \rightarrow \Z_{2}^{2^{s-1}}$ was discovered
by Carlet in \cite{car98}. Using this {\em generalized Gray map} he extendes
the concept of $\Z_4$-linear codes to $\Z_{2^s}$-linear codes. The generalization
of these isometries up to finite chain rings was considered by
Greferath and Schmidt, Honold and Landjev, San Ling and Blackford and 
Nechaev in \cite{gs99,hola99,lb02,neku96}. In \cite{bh01} Asch and Tilborg 
has studied the isometry between $\Z_{p^2}$ and $\Z_p^{p}$.  More
recently an isometry between codes over $\Z_{2^s}$ and codes
over $\Z_4$ has been introduced by Vega and Tapia-Recillas in
\cite{tv03,tv03dm}. Further they have studied the condition
under which a code over $\Z_4$ is $\Z_{2^s}$ linear.  In this
paper we will introduce a Gray isometry between $\Z_{2^s}$
and $\Z_{2^{s-1}}$.  It is shown that the generalized
Gray map introduced by Carlet is equivalent to the product of
$s-1$ such maps. It also covers the various generalizations
introduced earliear. Some basic properties of these maps
are studied.

\section{Preliminaries And Notations}

A {\em code} $\C,$ of length $n$, over $\;\Z_{p^s}$ is simply a
subset of $\Z_{p^s}^{n}$.  If in addition it is an additive
subgroup of $\Z_{p^s}^{n}$ we call $\C$ a {\em linear code} of
length $n$.  An element of $\C$ is called a {\em codeword of}
$\C$ and a {\em generator matrix} of a linear code $\C$ is a
matrix whose rows generate $\C$.  The {\em Hamming weight}
$w_H(x)$ of a vector $x$ in $\ZZ_{p^s}^n$ is the number of
non-zero components.  The {\em Lee weight} $w_L(x)$ of a
vector $x=(x_1,x_2,\ldots,x_n)$ is $\sum_{i=1}^n
\min\{|x_i|,|p^s-x_i|\}$.  The {\em Euclidean weight}
$w_E(x)$ of a vector $x$ is $\sum_{i=1}^n \min\{x_i^2,
(p^s-x_i)^2\}$.  The Euclidean weight is useful in connection
with lattice constructions.  The {\em Chinese Euclidean weight}
$w_{CE}(x)$ of a vector $x \in \Z_m^{n}$ is $\sum_{i=1}^n
\left\{2-2 \; \mbox{cos}\;\left(\frac{2 \pi x_i}{m}\right)\right\}$.
This is useful for $\!m\!-\!PSK$ coding.  In \cite{cohe97}, Constantinescu 
and Werner introduced the {\em homogeneous weight} on the residue class 
rings of integers which was later extended to finite Frobenius rings. 
The homogeneous weight on $\Z_{p^s}$ is defined as follows.  For $\; u  \neq 0, u \in \Z_{p^s}$
\[
w_{HW}(u)= \left\{
\begin{array}{cc} & p^{s-2}(p-1), u \notin p^{s-1}\Z_{p^s} \\
		   & p^{s-1}, u \in p^{s-1}\Z_{p^s} 
\end{array}\right. \;\mbox{and}\;w_{HW}(0)=0.
\]
\noindent The Hamming, Lee, Euclidean and Homogeneous distances $d_H(x,y)$, $d_L(x,y)$, 
$d_E(x,y)$ and $d_{HW}(x,y)$ between two vectors $x$ and $y$ are
$w_H(x-y)$, $w_L(x-y)$, $w_E(x-y)$ and $w_{HW}(x-y)$ respectively.  The minimum
Hamming, Lee, Euclidean and Homogeneous weights, $d_H, d_L$, $d_E$ and $d_{HW}$
of $\C$ are the smallest Hamming, Lee, Euclidean and Homogeneous weights
among all non-zero codewords of $\C,$ respectively.

Carlet's defenition of generalized Gray map can be extended
to $\Z_{p^s}$ naturally.  The {\em Generalized Gray map}
$\gamma : \;\Z_{p^s}^{n} \rightarrow \;\Z_p^{p^{s-1}n}$ is
the coordinate-wise extension of the function $\gamma_g$
from $\;\Z_{p^s}$ to $\;\Z_p^{p^{s-1}}$ defined as follows.
Let $u = u_1+p u_2 + \cdot + p^{s-1}u_s \in \Z_{p^s}$ be
the finite $p$-adic expansion of $u$, where $u_i \in \Z_p$
also let $y_i \in \Z_p$. The map $\gamma_g:(y_1,y_2,
\ldots, y_{s-1}) \mapsto u_s + \sum_{i=1}^{s-1}u_i y_i
\in \Z_p$ defined as a generalized Boolean function
evaluated on $GF(p^{s-1})$. It is easy to see that this
map is equivalent to the generalized Gray map introduced
by Greferath and Schmidt \cite{gs99} using tensor product.
In particular for $s=2$ this map is also equivalent
to a generalized Gray map from $\Z_{p^2}$ to $\Z_p^{p}$
introduced recently by Asch and Tilborg \cite{bh01}.  The image
 $\gamma(\C)$, of a linear code $\C$ over $\;\Z_{p^s}$
of length $n$ by the generalized Gray map, is a $p$-ary
code of length $p^{s-1}n$.  The {\em dual code} $\C^{\perp}$
 of $\C$ is defined as $\{x \in {\ZZ}_{p^s}^n \mid
x \cdot y = 0\; \mbox{for all} \; y\in \C\}$ where $x
\cdot y$ is the standard inner product of $x$ and $y$.  $\C$
is {\em self-orthogonal} if $\C \subseteq \C^\perp$ and $\C$
is {\em self-dual} if $\C=\C^\perp$.  Two codes are said
to be {\em equivalent} if one can be obtained from the other
by permuting the coordinates and (if necessary) changing the
signs of certain coordinates.  Codes differing by only a
permutation of coordinates are called {\em permutation-equivalent}.


Any linear code $\C$ over $\ZZ_{p^s}$ is permutation-equivalent
to a code with generator matrix $G$ (the rows of $G$
generate $\C$) of the form
\begin{equation}\label{eqn:01}
G=\left[\begin{array}{cccccc}
I_{k_0}&A_{01}&A_{02}& \cdots&A_{0s-1}&A_{0s}\\
{\bf 0}&pI_{k_1}&pA_{12}& \cdots &pA_{1s-1}& pA_{1s}\\
{\bf 0}&{\bf 0}&p^{2}I_{k_2}& \cdots & p^{2}A_{2s-1} & p^{2}A_{2s}\\
\vdots& \vdots & \vdots & \ddots & \vdots & \vdots \\
{\bf 0}&{\bf 0}&{\bf 0}& \cdots & p^{s-1}I_{k_{s-1}} & p^{s-1}A_{s-1s}
\end{array}\right],
\end{equation}
\noindent where $A_{ij}$ are matrices over
$\;\Z_{p^{s}}$ and the columns are grouped into blocks  of
sizes $k_0, \; k_1, \; \cdots, \; k_{s-1}, \; k_{s},$ respectively.
Let $k=\sum_{i=0}^{s-1} (s-i)k_i$. Then $|\C|= p^{k}$ is
the number of codewords in the code $\C.$ Note that $\C$ is
a free code if and only if $k_i=0 \;\;\mbox{for all}\; \;\;
i=1,2, \cdots, s-1.$ It is easy to see that a generator
matrix for $\C^{\perp}$ is of the form
\begin{center}
$
H=\left[\begin{array}{ccccc}B_{0s}&B_{0s-1}&
\cdots & B_{01}& I_{k_{s}}\\
pB_{1s}& pB_{1s-1}& \cdots &pI_{k_{s-1}}& {\bf 0}\\
\vdots & \vdots & \ddots & \vdots & \vdots \\
p^{s-1}B_{s-1s}&p^{s-1}I_{k_1}& \cdots & {\bf 0} & {\bf 0} \end{array}
\right],
$ \end{center}
\noindent where $B_{ij}$ are matrices over $\;\Z_{p^{s}}$
and the columns are grouped into blocks of size $k_0,k_1,
\cdots, k_s.$ Therefore $|\C^{\perp}|= p^{sn-k}.$ Note
that $sn-k=\sum_{i=1}^{s}ik_i.$

  A vector $\v$ is a {\em $p$-linear combination} of the
vectors $\v_1, \v_2, \ldots, \v_k$ if
$ \v = \l_1 \v_1 + \ldots + \l_k \v_k$ with
$\l_i \in \Z_p$ for $ 1 \leq i \leq k.$ A subset
$ S = \{ \v_1,\v_2, ...,\v_k \}$ of $\C$ is called a {\em $p$-basis}
for $\C$ if  for each $i= 1,2,...,k-1, \; p \v_i$ is a
$p-$linear combination of $ \v_{i+1},..., \v_k$, $ 2 \v_k = 0,
\; \C$ is the $p$-linear span of $S$ and $S$ is $p$-linearly
independent \cite{gu01}. The number of elements in a \
$p$-basis for $\C$ is called the {\em $p$-dimension} of $\C.$
      It is easy to verify that the rows of the matrix
\begin{equation}\label{eqn:02}
\B=\left[\begin{array}{cccccc}
I_{k_0}&A_{01}&A_{02}& \cdots&A_{0s-1}&A_{0s}\\
pI_{k_0}&pA_{01}&pA_{02}& \cdots&pA_{0s-1}&pA_{0s}\\
\vdots& \vdots & \vdots & \ddots & \vdots & \vdots \\
p^{s-1}I_{k_0}&p^{s-1}A_{01}&p^{s-1}A_{02}& \cdots&p^{s-1}A_{0s-1}&
p^{s-1}A_{0s}\\\hline
{\bf 0}&pI_{k_1}&pA_{12}& \cdots &pA_{1s-1}& pA_{1s}\\
{\bf 0}&p^{2}I_{k_1}&p^{2}A_{12}& \cdots &p^{2}A_{1s-1}& p^{2}A_{1s}\\
\vdots& \vdots & \vdots & \ddots & \vdots & \vdots \\
{\bf 0}&p^{s-1}I_{k_1}&p^{s-1}A_{12}& \cdots &p^{s-1}A_{1s-1}&
p^{s-1}A_{1s}\\\hline
\vdots& \vdots & \vdots & \ddots & \vdots & \vdots \\\hline
{\bf 0}&{\bf 0}&{\bf 0}& \cdots & p^{s-1}I_{k_{s-1}} & p^{s-1}A_{s-1s}
\end{array}\right].
\end{equation}
form a $p$-basis for the code $\C$
generated by $G$ given in (1).

A linear code $\C$ over $\;\Z_{p^s}$ ( over $\;\Z_2$) of length $n$,
$p$-dimension $k$, minimum distance $d_H, d_L, d_E, d_{CE}$
and $d_{HW}$ is called an $\left[ n,k,d_H,d_L,d_E,d_{CE},d_{HW}
\right]$ $\left([n,k,d_H]\right)$ or simply an
$\left[ n,k \right]$ code.

\section{Modular Gray Map}

Now we define a natural Gray map as follows.

Let
\[
\begin{array}{ccc}
A_1&=&\{0,1,2,3,\ldots,2^{s-2}-1\},\\
A_2&=&\{2^{s-2},2^{s-2}+1,\ldots,2^{s-1}-1\},\\
A_3&=&\{2^{s-1},2^{s-1}+1,\ldots,2^{s-1}+2^{s-2}-1\}\;\mbox{and}\\
A_4&=&\{2^{s-1}+2^{s-2},2^{s-1}+2^{s-2}+1,\ldots,2^{s-1}+2^{s-1}-1\}
\end{array}
\]
\noindent be the four consecutive disjoint partitions of the
integers of $\Z_{2^s}$. Define the modular generalized Gray map
$\eta^s_g$ on $\Z_{2^s}$ to $\Z_{2^{s-1}}$ as

\[
\eta^s_g(i)=
\left\{
\begin{array}{cc} (i,i) & i \in A_1\\
(i-2^{s-2},i) & i \in A_2\\
(i-2^{s-2},i-2^{s-2}) & i \in A_3\\
(i-2^{s-1},i-3 \cdot 2^{s-2}) & i \in A_4
\end{array}\right.
\]

\begin{remark}\label{rm}
Each entry of $\Z_{2^s}$ is mapped onto the first order Reed
Muller code of length $2$ over $\Z_{2^s}$ \cite{gu01,gbl05} generated by the
following generator matrix

\[
G=\left[\begin{array}{cc} 1& 1 \\
0 & 2^{s-2}
\end{array}\right]_{2 \times 2}.
\]
\end{remark}

\begin{remark}
For $s=2$ $\eta^s_g$ reduces to ordinary Gray map.
\end{remark}

\begin{remark}
Applying $\eta^s_g$ to a code $\C$ over $\Z_{2^s}$ gives a code
$\eta^s_g(C)$ over $\Z_{2^{s-1}}$ of length twice the length of
$\C$
\end{remark}

\begin{example}
For $s=2,3,4$ these maps are given below.
\end{example}

\[
\begin{array}{ccc}
\Z_4 & \mapsto & \Z_2 \\
0 & \mapsto & 00 \\
1 & \mapsto & 01 \\
2 & \mapsto & 11 \\
3 & \mapsto & 10
\end{array}
\]

\[
\begin{array}{ccc}
\Z_8 & \mapsto & \Z_4 \\
0 & \mapsto & 00 \\
1 & \mapsto & 11 \\
2 & \mapsto & 02 \\
3 & \mapsto & 13 \\
4 & \mapsto & 22 \\
5 & \mapsto & 33 \\
6 & \mapsto & 20 \\
7 & \mapsto & 31
\end{array}
\]

\[
\begin{array}{ccc}
\Z_{16} & \mapsto & \Z_8 \\
0 & \mapsto & 00 \\
1 & \mapsto & 11 \\
2 & \mapsto & 22 \\
3 & \mapsto & 33 \\
4 & \mapsto & 04 \\
5 & \mapsto & 15 \\
6 & \mapsto & 26 \\
7 & \mapsto & 37 \\
8 & \mapsto & 44 \\
9 & \mapsto & 55 \\
10 & \mapsto & 66 \\
11 & \mapsto & 77 \\
12 & \mapsto & 40 \\
13 & \mapsto & 51 \\
14 & \mapsto & 62 \\
15 & \mapsto & 73
\end{array}
\]

\noindent Some interesting facts about modular Gray maps are collected in the following.

\begin{theorem}
The generalized Gray map of Carlet $\Phi_G : \Z_{2^s} \mapsto
\Z_2$ can written as a product of $s-1$ modular Gray maps i.e,
$\Phi_G = \eta^s_g \eta^{s-1}_g\eta^{s-2}_g\ldots \eta^2_g$.
\end{theorem}

\proof By remark \ref{rm}, if we apply the modular gray map $\eta^{s}_g$ to the $2$-basis 
generator matrix of the first order Reed Muller code over $\Z_{2^s}$ we get the $2$-basis 
generator matrix of the first order Reed Muller code over $\Z_{2^{s-1}}$ which is the image set of 
$\eta^{s-1}_g$ and so on. Thus successive applications yields  Carlet's generalized Gray map 
\cite{gu01,gbl05}.

\qed

\begin{remark}
It is interesting to note that now we can get modular Gray map between all intermediate rings i.e, 
between 
\[
\Z_{2^s} \mapsto \Z_{2^{s-1}} \mapsto \Z_{2^{s-2}} \mapsto \cdots \mapsto \Z_4 \mapsto \Z_2.
\]
For example, the Gray map of Tapia-Recillas and Vega \cite{tv03,tv03dm} between $\Z_{2^s}$ and $\Z_4$ is equivalent to a map by taking product of $s-2$ such maps $\eta^s_g \eta^{s-1}_g \ldots \eta^{3}_g$. Clearly our map
is much more powerful.

\end{remark}


\begin{lemma}\label{lemim}
The images of the rows of the matrix $\B$ given by (\ref{eqn:02})
under the modular Gray map $\eta^s_g$ are $2$-linearly independent
over $\;\Z_{2^{s-1}}.$
\end{lemma}

\proof  After taking appropriate block permutations row-wise to the matrix $\B$ and applying the modular 
Gray map we see that in the diagonals we have 2-linearly independent vectors over $\Z_{2^{s-1}}$.  This proves our claim.

\qed

\begin{proposition}
The modular Gray map $\eta^s_g$ is an isometry between $\Z_{2^s} \mapsto (\Z_{2^{s-1}})^2$ 
with respect to the homogeneous weights.
\end{proposition}

\proof Let $u (\neq 0) \in \Z_{2^s}$.  Then the homogeneous weight will be given as 
\[
w_{HW}(u)=\left\{\begin{array}{cc}2^{s-2},&u \neq 2^{s-1}\\
2^{s-1},& u=2^{s-1}.\end{array}\right.
\] Thus homogeneous weight of each entry in the sets $A_1, A_2, A_3\;(\mbox{except first entry which is}\; 2^{s-1})$ and $A_4$ will be $2^{s-2}$. Thus direct computation of homogeneous weight as a rational sum of the entries in the definition yields the desired result. The case of $u=0$ is trivial.
\qed

\begin{remark}
Note that any product of modular Gray maps $\eta^s_g$ is also an isometry. 
\end{remark}

\begin{remark}
The modular Gray map $\eta^s_g$ can be extended coordinate-wise 
to a map from $(\Z_{2^s})^n$ to $(\Z_{2^{s-1}})^{2n}$. With abuse of notation we 
denote this map also by $\eta^s_g$. 
\end{remark}

\subsection{Permuted Modular Gray Map}
We can permute the modular Gray map to get some very interesting
results like perverseness of Octocode. The modified modular Gray
map $\xi^s_g$ is define as ....

....

Let us modify our examples to get the modify modular Gray map for
$s=2,3,4$.

\[
\begin{array}{ccc}
\Z_4 & \mapsto & \Z_2 \\
0 & \mapsto & 00 \\
1 & \mapsto & 10 \\
2 & \mapsto & 11 \\
3 & \mapsto & 01
\end{array}
\]

\[
\begin{array}{ccc}
\Z_8 & \mapsto & \Z_4 \\
0 & \mapsto & 00 \\
1 & \mapsto & 11 \\
2 & \mapsto & 20 \\
3 & \mapsto & 13 \\
4 & \mapsto & 22 \\
5 & \mapsto & 31 \\
6 & \mapsto & 02 \\
7 & \mapsto & 33
\end{array}
\]

\[
\begin{array}{ccc}
\Z_{16} & \mapsto & \Z_8 \\
0 & \mapsto & 00 \\
1 & \mapsto & 11 \\
2 & \mapsto & 22 \\
3 & \mapsto & 13 \\
4 & \mapsto & 40 \\
5 & \mapsto & 15 \\
6 & \mapsto & 26 \\
7 & \mapsto & 17 \\
8 & \mapsto & 44 \\
9 & \mapsto & 71 \\
10 & \mapsto & 62 \\
11 & \mapsto & 73 \\
12 & \mapsto & 04 \\
13 & \mapsto & 75 \\
14 & \mapsto & 66 \\
15 & \mapsto & 77
\end{array}
\]

\begin{example}
Consider the generator matrix of the Octocode over $\Z_8$.
\[
G_8=\left[\begin{array}{cccccccc}
5 & 7 & 5 & 6 & 1 & 0 & 0 & 0 \\
5 & 0 & 7 & 5 & 6 & 1 & 0 & 0 \\
5 & 0 & 0 & 7 & 5 & 6 & 1 & 0 \\
5 & 0 & 0 & 0 & 7 & 5 & 6 & 1
\end{array}
\right].
\]
\noindent Apply the modular Gray map $\eta^3_g$ on the rows of
$G_8$. We get the following generator matrix 
\[
\eta^3_g(G_8) = \left[\begin{array}{cccccccc|cccccccc}
3 & 3 & 3 & 0 & 1 & 0 & 0& 0 & 1 & 3 & 1 & 2 & 1 & 0 & 0 & 0 \\
3 & 0 & 3 & 3 & 0 & 1 & 0& 0 & 1 & 0 & 3 & 1 & 2 & 1 & 0 & 0 \\
3 & 0 & 0 & 3 & 3 & 0 & 1& 0 & 1 & 0 & 0 & 3 & 1 & 2 & 1 & 0 \\
3 & 0 & 0 & 0 & 3 & 3 & 0& 1 & 1 & 0 & 0 & 0 & 3 & 1 & 2 & 1
\end{array}
\right].
\]
\end{example}

\section{Properties}
In this section we list several interesting properties of modular Gray map that one can study.

\begin{itemize}
\item General results of Theorems $4,5,6,$ page $305$ of \cite{hkcss94} these should also
generalize the results of Vega et al in \cite{tv03} and Carlet's result of \cite{car98}. 
\item Studying the characterizations of cyclic, quasi-cyclic, consta-cyclic, negacyclic, quasi-twisted codes with 
respect to modular Gray map in view of the following papers \cite{wo99,wo01,lb02,tv03dm,vw04}.
\item Does modular Gray map preserves self orthogonality of codes ? Under what conditions ?
\item Applications of both modular and permuted modular Gray map for the constructions of new codes 
over various modular rings and binary fields. For example to Hamming codes, Simplex codes, Golay codes and other standard codes.
\item  Is there any connection for the modular Gray map with the Nenchev's Gray map ?
\item Other non-trivial applications like by taking the generator matrix of Kerdock codes, Preparata codes, 
Goethals codes or like codes and applying the modular Gray map.
\item Construction of record breaking codes !
\item Effect on Torsion and reduction codes !
\item New code construction with respect to 2-basis modular Gray map applications !
\item Connection with Hensal-uplift !
\item Studying the $\Z_{2^s}$-linearity of standard uplifted codes over $\Z_{2^{s-1}}$.
\item Finding other non-trivial properties ! 
\item ...

\end{itemize}

{\bf Note}
This note is an old contribution and has not been updated with some new results on Gray map. We refer the reader to look into some new literature in the area. 


\end{document}